# A RESAMPLING-BASED TEST TO DETECT PERSON-TO-PERSON TRANSMISSION OF INFECTIOUS DISEASE[1]

By Yang Yang, Ira M. Longini, Jr. and M. Elizabeth Halloran

*Fred Hutchinson Cancer Research Center, Fred Hutchinson Cancer Research Center and University of Washington, Fred Hutchinson Cancer Research Center and University of Washington*

Early detection of person-to-person transmission of emerging infectious diseases such as avian influenza is crucial for containing pandemics. We developed a simple permutation test and its refined version for this purpose. A simulation study shows that the refined permutation test is as powerful as or outcompetes the conventional test built on asymptotic theory, especially when the sample size is small. In addition, our resampling methods can be applied to a broad range of problems where an asymptotic test is not available or fails. We also found that decent statistical power could be attained with just a small number of cases, if the disease is moderately transmissible between humans.

**1. Introduction.** Most emerging infectious disease pathogens in humans cross from their natural zoonotic reservoir to human populations where early mutated, reassorted or recombined forms begin to spread from person-to-person [Antia et al. (2003)]. Examples include human immunodeficiency virus, monkey pox, severe acute respiratory syndrome and pandemic influenza. Currently, a highly pathogenic avian influenza strain (H5N1) has been spreading from poultry to humans, mostly in Southeast Asia, with possible limited human-to-human spread through close contact in Indonesia [Butler (2006)]. A concern is that this virus could cause a large scale pandemic as it becomes more adapted to human-to-human transmission. Real-time surveillance provides limited information on small clusters of human cases in terms of symptom onset times and physical location. It is critical to

Received November 2006; revised February 2007.

[1]Supported by the National Institute of General Medical Sciences MIDAS Grant U01-GM070749 and the National Institute of Allergy and Infectious Diseases Grant R01-AI32042.

*Key words and phrases.* Resampling, permutation, hypothesis test, infectious disease, transmission, likelihood ratio, MLE.

answer two questions in real time: 1. Is the infectious agent spreading from person to person? and 2. If it is, how transmissible is it? The first question is novel and, to our knowledge, has not been addressed in the statistical literature. The second question is an estimation problem, and various statistical methods using household data are applicable, such as the models based on observed final infection status [Longini and Koopman ([1982](#)), Becker and Hasofer ([1997](#)), O'Neill and Roberts ([1999](#))] and those based on a discrete-time sequence of symptom onset [Rampey et al. ([1992](#)), Yang, Longini and Halloran ([2006](#))]. Our major goal in this paper is to answer the first question, but an estimation method is needed for this goal. We base our approach on that in Yang, Longini and Halloran ([2006](#)).

The statistical questions hinge on inference about the transmissibility of the infectious agent. The basic reproductive number, $R_0$, is the fundamental measure of the transmissibility of an emerging infectious agent. Given that the emerging infectious agent is transmissible, estimates of $R_0$ will generally be small and are not very informative. In addition, estimation of some epidemic characteristics such as secondary attack rates (SAR) and $R_0$ heavily relies on the specification of a correct transmission model. When there is no person-to-person transmission, estimates of these characteristics may be nonzero, but are not meaningful. Therefore, a test of the existence of person-to-person transmission can provide a solid ground for parameter estimation. Specifically, one would like to test whether the person-to-person transmission probability, no matter how it is defined, is 0. As a probability always takes values from 0 to 1, the boundary value 0, which is a nonstandard condition, imposes an immediate challenge, because the null distribution of standard statistics, based on which tests are performed, are generally difficult to track. Although statisticians have discussed asymptotic tests for a limited set of scenarios [Moran ([1971](#)), Self and Liang ([1987](#)), Feng and McCulloch ([1992](#))], more often such an asymptotic null distribution is not available for a specific case. Furthermore, the validity of asymptotic tests depends on relatively large sample sizes, which may compromise the power of such tests to detect person-to-person transmission if applied to a small sample size, such as those generated by avian influenza. These challenges motivate our investigation in exact rather than asymptotic testing methods.

**2. Methods.** The data structure we usually observe is a sequence of symptom onsets and associated cluster information, for example, at what time a symptom onset occurred in which household. To construct a probability model with a reasonable level of complexity from the observed data, it is necessary to make basic assumptions about the natural history of the disease and the transmission mechanism. We assume that the incubation period is the same as the latent period, but other assumptions could be made about the relation of the two periods. We make the following additional assumptions about the disease. Any newly infected person remains asymptomatic over a period of $\delta$ days (the incubation period) before symptom onset, where $\delta$ is a random quantity with a distribution of $g(l) = \Pr(\delta = 1)$, $l = \delta_{\min}, \delta_{\min} + 1, \ldots, \delta_{\max}$. We denote by $\delta_{\min}$ and $\delta_{\max}$ the minimum and maximum durations (in days) of the latent period. Upon symptom onset, the person becomes and remains infectious over a period of $\eta$ days (infectious period), where $\eta$ is also a random quantity with a distribution $f(l) = \Pr(\eta = l)$, $l = \eta_{\min}, \eta_{\min} + 1, \ldots, \eta_{\max}$. Similarly, $\eta_{\min}$ and $\eta_{\max}$ are the minimum and maximum durations of the infectious period. In this paper our method requires pre-specifying $g(l)$ and $f(l)$.

We consider the dynamic of a community-based epidemic on a day-by-day basis. We assume that the whole community is exposed to some external source with a constant level of infectivity for $S$ days. Such an external common source takes into account all possible channels, such as exposure to infected animals, through which the disease can be introduced into the community. Let $b$ be the probability that a susceptible person in the community is infected by the common source during one day of exposure. The probability of infection by the common source throughout the $S$-day exposure period is called the community probability of infection (CPI) and is given by $1 - (1 - b)^S$ [Longini and Koopman ([1982](#))]. Once the disease is introduced into the community, transmission between people may occur through contacts. There are various types of contacts one can define. We define a contact as all possible interactions during one day that can potentially transmit the disease from an infective person to a susceptible person. We consider two levels of contacts: close contacts between two persons who live in the same household and casual contacts between two persons who live in different households but may make contact in the community. We denote by $p_1$ the daily probability of transmission with a close contact, and by $p_2$ with a casual contact.

With the above setting, we can construct a likelihood and obtain the maximum likelihood estimates (MLEs) for the unknown parameters ($b$, $p_1$ and $p_2$) as given in the [Appendix](#). Two quantities related to transmission probabilities that we would also like to estimate are the SAR and $R_0$. The SAR is defined as the probability of infection if a susceptible is exposed to an infective during his or her infectious period. Corresponding to the levels of contact, there are two types of SAR defined as $\text{SAR}_k = \sum_l f(l)(1 - (1 - p_k)^l)$, $k = 1, 2$. $\text{SAR}_1$ is the SAR within households and is of more epidemiological interest than $\text{SAR}_2$. The basic reproductive number refers to the expected number of people a typical infective person can infect among a large susceptible population. Here we are interested in the expected number of people that an infective person can infect given that he or she is the first infected person in this community. We refer to this as the local reproductive number $R$. Estimates of the local $R$ cannot be generalized to a broader context because of the potential selection bias. The clusters are often selected based on a number of cases and may represent higher $R_0$ than in the general population. For a community of $N$ households with a uniform household size $M$, we have $R = (M - 1) \times \mathrm{SAR}_1 + (N - M) \times \mathrm{SAR}_2$.

Nonzero estimates of $p_1$ or $p_2$ do not necessarily imply that their true values are nonzero. In addition, construction of valid 95% confidence intervals for the estimates of transmission probabilities is difficult when their true values are 0's. Therefore, a valid test of the hypothesis $p_1 = p_2 = 0$ would be of great public health interest. A formal statement of the hypothesis test is

$$\mathcal{H}_0 : p_1 = p_2 = 0 \quad \text{vs.}$$

$$\mathcal{H}_1 : p_1 > 0 \quad \text{or} \quad p_2 > 0,$$

where $\mathcal{H}_0$ is the null hypothesis and $\mathcal{H}_1$ is the alternative hypothesis.

A natural choice of test statistic is the likelihood ratio statistic

$$(1) \qquad \lambda = -2 \log \frac{\sup_b L_0(b|\tilde{t}_i, i = 1, \ldots, N)}{\sup_{b,p_1,p_2} L(b, p_1, p_2|\tilde{t}_i, i = 1, \ldots, N)},$$

where the numerator is the maximum likelihood (ML) when we restrict $p_1 = p_2 = 0$, and the denominator is the ML without such restriction, both conditioning on observed symptom onset times $\tilde{t}_i$ ($\tilde{t}_i = \infty$ for uninfected individuals). Explicit expression of the likelihoods are given in the Appendix. The likelihood ratio statistic asymptotically follows a Chi-square distribution with 2 degrees of freedom when $\mathcal{H}_0$ is true, if all regularity conditions hold for this probability structure [Lehmann (1999)]. However, two nonstandard conditions are present in our case. One is that the hypothesized parameter values under testing are boundary. As mentioned before, the asymptotic null distribution is generally difficult to track when boundary values are to be tested. Self and Liang (1987) discussed asymptotic distributions of the likelihood ratio statistic for some settings of boundary parameters, but our case is not one of them. The other nonstandard condition is that the parameters to be tested affect the domain of observable data. When $p_1 = p_2 = 0$, infections are confined to the $S$ days with exposure to the common infective source. Therefore, no symptom onset can happen after day $S + \delta_{\max}$. When $p_1 \neq 0$ or $p_2 \neq 0$, the domain of the observable data is much larger. No valid asymptotic test exists when this nonstandard condition is present, unless we only use the data up to day $S$ for testing at the price of losing some information.

Resampling methods have been widely applied to hypothesis testing, especially in the recent decade because of their easy implementation with modern computational capacity. While employing less stringent model assumptions, these methods can attain the same level of statistical power as standard tests [Hoeffding (1952), Box and Andersen (1955)]. Permutation tests (or randomization tests) have been well developed in the setting of two-sample comparison and ANOVA [Fisher ([1935](#)), Pitman ([1937](#)), Welch ([1990](#))]. For the boundary problem with parameter values specified by $\mathcal{H}_0$, the bootstrap was used in combination with the likelihood ratio statistic to test the number of components in mixture models [McLachlan ([1987](#)), Feng and McCulloch ([1996](#))]. We propose two approaches, a simple permutation test and a more refined one, for the problem of testing the person-to-person transmission probability. These resampling-based methods do not suffer from the two nonstandard conditions mentioned above, as shown by a simulation study. When the observed data are truly generated from $\mathcal{H}_0$, we can reassign all of the observed symptom onset days (and associated infection status) to a different collection of individuals, and every such rearrangement is equally likely with the same likelihood $L_0$. The empirical distribution of the test statistic calculated from permuting symptom onset days across the population can then be used to approximate the null distribution under $\mathcal{H}_0$. This simple permutation test can be refined by varying symptom onset days of infected individuals in any given permuted data while keeping the likelihood $L_0$ under the null hypothesis unchanged. The refined permutation test resamples data points from a much larger sampling space as compared to the simple permutation test. Technical details concerning development of the two resampling methods can be found in the [Appendix](#).

We first use simulations to verify the validity of the resampling methods by comparing them to the asymptotic test for a simpler scenario with only $b$ and $p_1$, that is, person-to-person transmission can only happen within households. For this two-parameter setting, Self and Liang ([1987](#)) showed that $\lambda$ will asymptotically follow a mixture distribution of $\chi_0^2$ and $\chi_1^2$ with equal mixing probability. Only data up to day $S$ are used for such comparison with the asymptotic test. We found that the refined permutation test has the best performance in terms of preserving type I error at the pre-specified level and yielding higher statistical power when population size and the number of cases are small. Results and discussion for the simple scenario are provided in the [Appendix](#) as well. Then we use simulations to investigate the performance of the refined permutation test for the scenario with three parameters: $b$, $p_1$ and $p_2$.

Computing $\lambda$ involves calculating likelihoods under two different models, the one with restriction of parameters conforming to $\mathcal{H}_0$ is the null model, and the other one without any restriction is the full model. For a realized epidemic, one of the two models may not be admissible (or possible). For example, when the minimum interval between any pair of consecutive cases is larger than the maximum duration of the latent period, no infection can be possibly attributed to person-to-person infection; thus, only the null model is admissible. On the other hand, when there is any case on or after the day $S + \delta_{\max}$, where $\delta_{\max}$ is the maximum duration of the latent period, only the full model is admissible because the common source is infective up to day $S$. When only the null (full) model is admissible, the p-value for that epidemic is assigned 1 (0). Resampling-based tests are performed only when both models are admissible. Checking admissibility can help avoid nonconvergence problems when maximizing likelihoods.

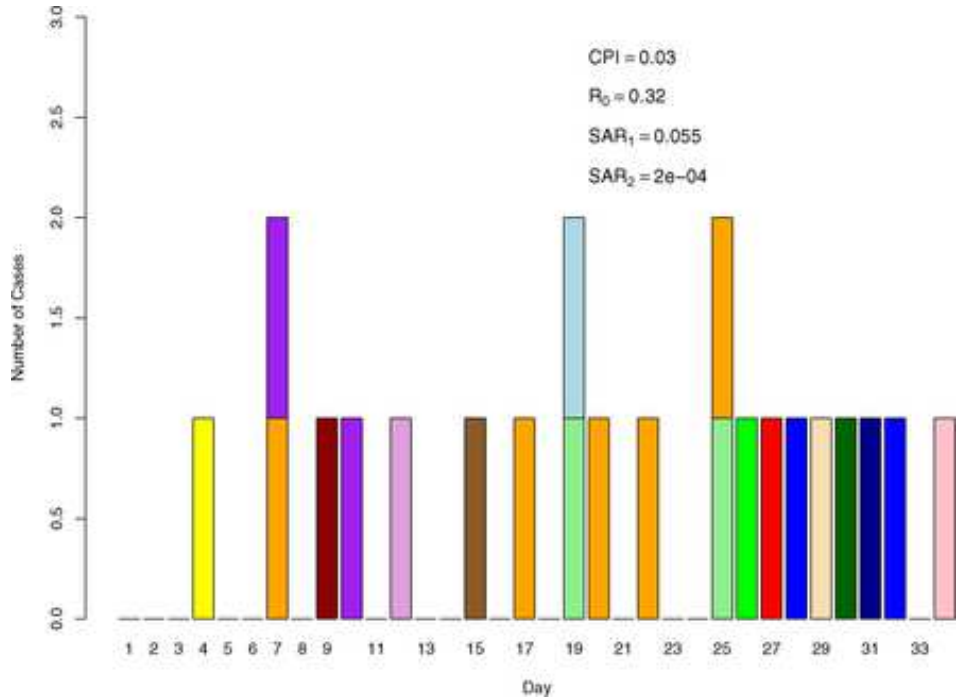

Fig. 1. *A sample epidemic curve for* $b = 0.001$, $p_1 = 0.014$ *and* $p_2 = 0.00005$. *Cases from the same household have the same color.*

**3. Results.**   For simplicity, we simulate epidemics over a community composed of 100 households, each of size 5. We let the exposure to external common source last $S = 30$ days, and let the epidemic exhaust itself. We do not introduce initial cases to start the epidemic, but let the common source initiate infection. Simulation runs with zero infections were discarded. We simulate epidemics based on $g(l) = \frac{1}{3}$, $l = 1, 2, 3$, and $f(l) = \frac{1}{3}$, $l = 3, 4, 5$, and these distribution are correctly specified by the methods that we evaluate. All p-values presented in this section are obtained by the refined permutation test, but simulations show that the simple permutation method gives similar results under the same population and parameter settings as discussed here, except that it tends to be too conservative about preserving type I error for extremely small $b$.

Table 1

*Power ($\times 100$) to detect person-to-person transmission for different settings of $b$ and $p_1$, with $p_2$ fixed at 0.00005 ($SAR_2 = 0.0002$). Numbers in parentheses are the average number of index cases over the average total number of cases. Results are based on 2000 simulations. 2000 permuted samples were drawn for each permutation test*

| $p_1$ | $b$ | | | | | | | | | | $SAR_1$ | $R$ |
| | 0.0002 | 0.0004 | 0.0006 | 0.0008 | 0.0010 | 0.0012 | 0.0014 | 0.0016 | 0.0018 | 0.0020 | | |
|---|---|---|---|---|---|---|---|---|---|---|---|---|
| $0.0^a$ | $4.3(\frac{3}{3})$ | $4.8(\frac{6}{6})$ | $5.0(\frac{9}{9})$ | $5.1(\frac{11}{12})$ | $5.3(\frac{14}{15})$ | $4.2(\frac{17}{18})$ | $4.8(\frac{19}{21})$ | $4.9(\frac{21}{23})$ | $4.9(\frac{24}{26})$ | $5.0(\frac{26}{29})$ | 0.0 | 0.0 |
| 0.004 | | | | | | | | $62(\frac{24}{28})$ | $62(\frac{26}{31})$ | $67(\frac{29}{34})$ | 0.016 | 0.16 |
| 0.006 | | | | | $62(\frac{15}{18})$ | $68(\frac{18}{21})$ | $71(\frac{21}{25})$ | $72(\frac{24}{29})$ | $75(\frac{26}{32})$ | $79(\frac{29}{36})$ | 0.024 | 0.19 |
| 0.008 | | | | $66(\frac{13}{15})$ | $75(\frac{15}{19})$ | $79(\frac{18}{22})$ | $81(\frac{21}{26})$ | $84(\frac{24}{30})$ | $84(\frac{26}{33})$ | $87(\frac{28}{36})$ | 0.032 | 0.23 |
| 0.010 | | | $68(\frac{10}{12})$ | $75(\frac{12}{16})$ | $80(\frac{15}{19})$ | $84(\frac{18}{23})$ | $87(\frac{21}{27})$ | $90(\frac{24}{31})$ | $92(\frac{26}{34})$ | $92(\frac{29}{38})$ | 0.039 | 0.26 |
| 0.012 | | | $75(\frac{10}{12})$ | $81(\frac{13}{16})$ | $85(\frac{16}{20})$ | $90(\frac{18}{24})$ | $91(\frac{21}{28})$ | $95(\frac{24}{32})$ | $95(\frac{26}{36})$ | $96(\frac{29}{39})$ | 0.047 | 0.29 |
| 0.014 | | $72(\frac{7}{8})$ | $81(\frac{10}{13})$ | $87(\frac{13}{17})$ | $91(\frac{16}{21})$ | $93(\frac{19}{25})$ | $95(\frac{21}{29})$ | | | | 0.055 | 0.32 |
| 0.016 | | $77(\frac{7}{9})$ | $84(\frac{10}{13})$ | $90(\frac{13}{17})$ | | | | | | | 0.062 | 0.35 |
| 0.018 | | $78(\frac{7}{9})$ | $87(\frac{10}{14})$ | | | | | | | | 0.070 | 0.38 |
| 0.022 | | $85(\frac{7}{10})$ | | | | | | | | | 0.085 | 0.44 |
| 0.026 | | $88(\frac{7}{11})$ | | | | | | | | | 0.10 | 0.50 |
| 0.030 | $75(\frac{4}{6})$ | $92(\frac{7}{11})$ | | | | | | | | | 0.11 | 0.56 |
| 0.034 | $77(\frac{4}{6})$ | | | | | | | | | | 0.13 | 0.61 |
| 0.038 | $80(\frac{4}{7})$ | | | | | | | | | | 0.14 | 0.67 |
| 0.042 | $84(\frac{4}{8})$ | | | | | | | | | | 0.16 | 0.73 |
| 0.046 | $86(\frac{4}{8})$ | | | | | | | | | | 0.17 | 0.78 |
| CPI | 0.006 | 0.012 | 0.018 | 0.024 | 0.030 | 0.035 | 0.041 | 0.047 | 0.053 | 0.058 | | |

$^a$The presented values are type I errors when $p_1 = p_2 = 0.0$.

As $p_2$ is of limited interest, we fix it at 0.00005 ($SAR_2 = 0.0002$), and vary $b$ from 0.0002 to 0.002 (CPI from 0.006 to 0.058) with a step of 0.0002. We vary $p_1$ from 0.004 to 0.046 ($SAR_1$ from 0.016 to 0.17) with steps chosen specific to $b$ so as to yield power values in the range of $(0.6, 1.0)$. All tests are performed at the level of 0.05, that is, we intend to have type I errors of no more than 5% when $p_1 = p_2 = 0$. An epidemic curve of a sample run for $b = 0.001$ (CPI $= 0.03$) and $p_1 = 0.014$ ($SAR_1 = 0.055$) is displayed in Figure 1, with each block representing a symptomatic case. Cases from the same household are filled with the same color. A pattern is evident that cases in the same household tend to cluster together in time. The CPI, $R$ and SAR given in the figure are based on the true parameters, but they could be estimated from the data as well. Results based on 2000 simulations and 2000 permutations for each test are presented in Table 1. The first row where $p_1 = p_2 = 0$ gives type I errors for various values of $b$, from which it is observed that type I errors are all preserved at the specified level. As expected, larger $p_1$ yields higher power for fixed $b$; similarly, larger $b$ also yields higher power for any given $p_1$. Surprisingly, when there are as few as a total of only seven cases, it is still possible to have 80% power with a moderate $p_1$ ($SAR_1 = 0.14$), which means that person-to-person transmission can still be detected even when there is a very limited number of cases. This finding could be very useful as most avian influenza epidemics in humans in recent years have a scale of eight total cases or fewer. Of interest as well is that all of the $R$ values are below 1, as seen from the last column of Table 1.

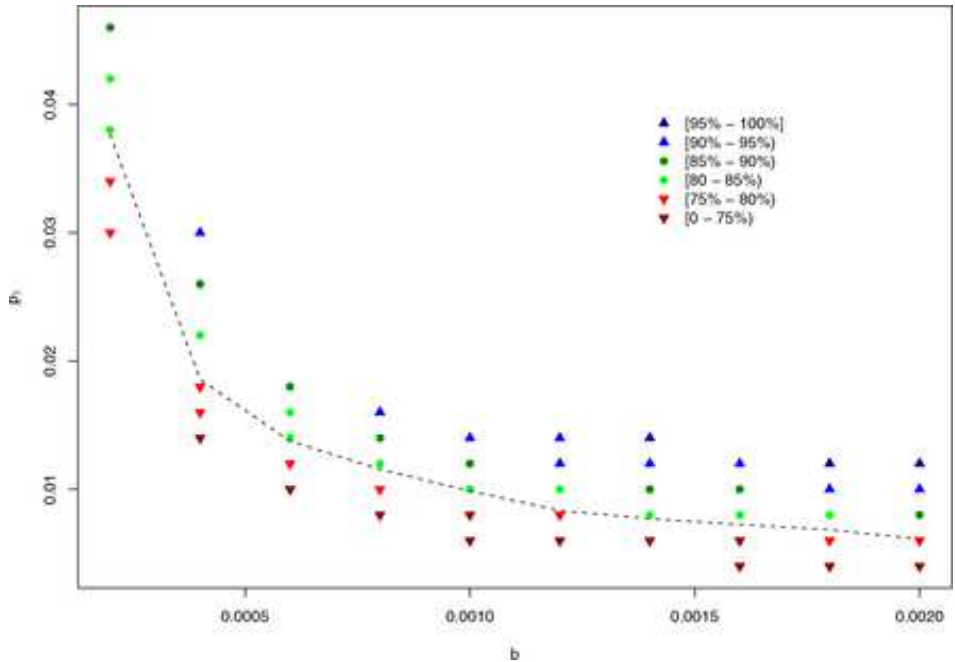

Fig. 2.  *Power to detect person-to-person transmission for different settings of $b$ and $p_1$, with $p_2$ fixed at 0.00005. Results are based on 2000 simulations. 2000 permuted samples were drawn for each permutation test. The dashed line is the 80% power contour line obtained from Loess smoothing.*

Figure 2 illustrates the information in Table 1, where power levels are shown in different colors and symbols with $b$ and $p_1$ as the horizontal and vertical axes, respectively. The 80% power contour curve obtained by Loess smoothing lies between green circles and red downward triangles. This figure clearly displays the trend of such a contour curve, descending sharply at $b = 0.0002$ ($CPI = 0.006$) and becoming flat around $p_1 = 0.008$ ($SAR_1 = 0.032$) as $b$ increases to 0.0014 ($CPI = 0.041$). Let $N_{idx}$ denote the mean number of index cases and $N_{tot}$ the mean total number of cases, averaging over all simulated epidemics. As only the number of cases are observable in real epidemics, we replace $b$ and $p_1$ with $N_{idx}$ and $N_{tot}$ as the axes in Figure 3. Not surprisingly, the underlying 80% power contour curve looks more linear, since roughly $N_{tot} \approx (1 + R)N_{idx}$. While $R$ depends on $p_1$, the range of $1 + R$ is relatively narrow, about $[1.2, 1.3]$ at $b \geq 0.0006$, and becomes narrower as $b$ increases. The figure also indicates that the power to detect person-to-person transmission is jointly determined by $N_{idx}$ and $N_{tot}$, instead of either alone. We fitted a linear regression between the complementary log–log transformed power values and selected transformations of $N_{idx}$ and $N_{tot}$, and found the following empirical formula:

$$\text{Power} = \exp\{-\exp(1.29 + 0.75N_{idx} - 0.55N_{tot} - 1.40\log(N_{idx}))\},$$

which explains 99% of the variation in power. Figure 4 plots the simulated vs. fitted power values, where most points fall close to the diagonal line, indicating that the empirical formula gives decent prediction, except for one point at $b = 0.0002$ and $p_1 = 0.03$, where the predicted power, 0.71, is somewhat lower than the simulated power, 0.75. Such an empirical formula could be used to predict power levels at various values of $N_{tot}$ and $N_{idx}$ for which simulations are not performed. The coefficients in the empirical formula will likely change for different parameter settings, and the linearity may not always hold.

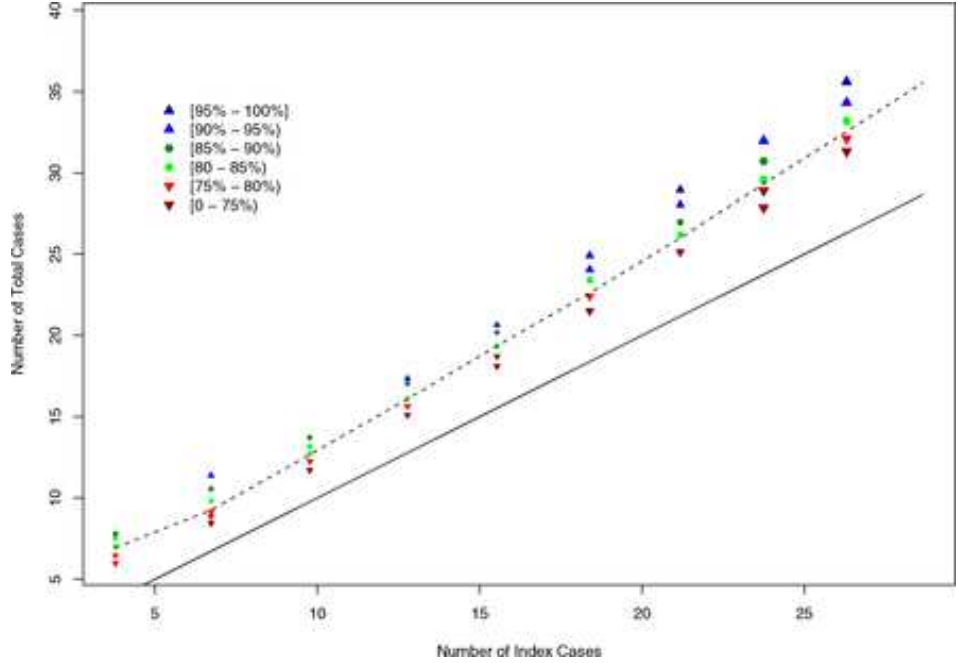

Fig. 3. *Power to detect person-to-person transmission plotted by the number of index cases versus the total number of cases. Results are based on 2000 simulations. 2000 permuted samples were drawn for each permutation test. The dashed line is the 80% power contour line obtained from Loess smoothing. The solid line is the lower bound (0) of power, where the number of index cases equals the total number of cases.*

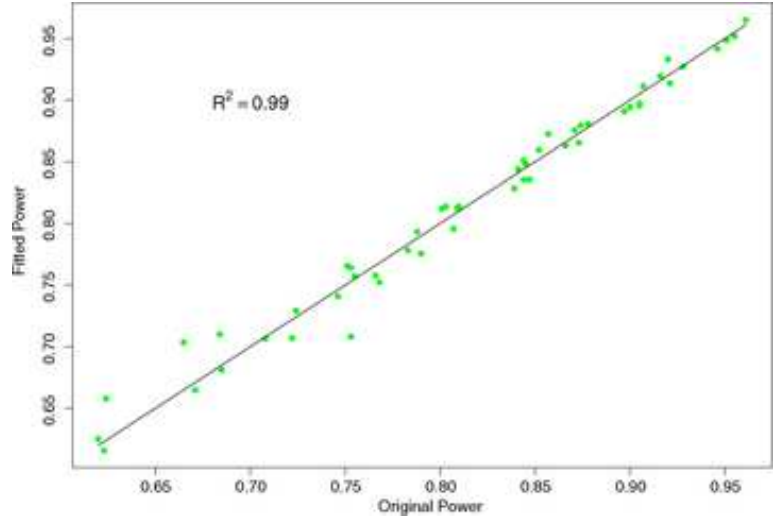

Fig. 4.  *Plot of simulated and fitted values of power from the empirical formula*
$\text{Power} = \exp\{-\exp(1.29 + 0.75N_{\text{idx}} - 0.55N_{\text{tot}} - 1.40\log(N_{\text{idx}}))\}$. *A good formula should
have all the points falling close to the diagonal line.*

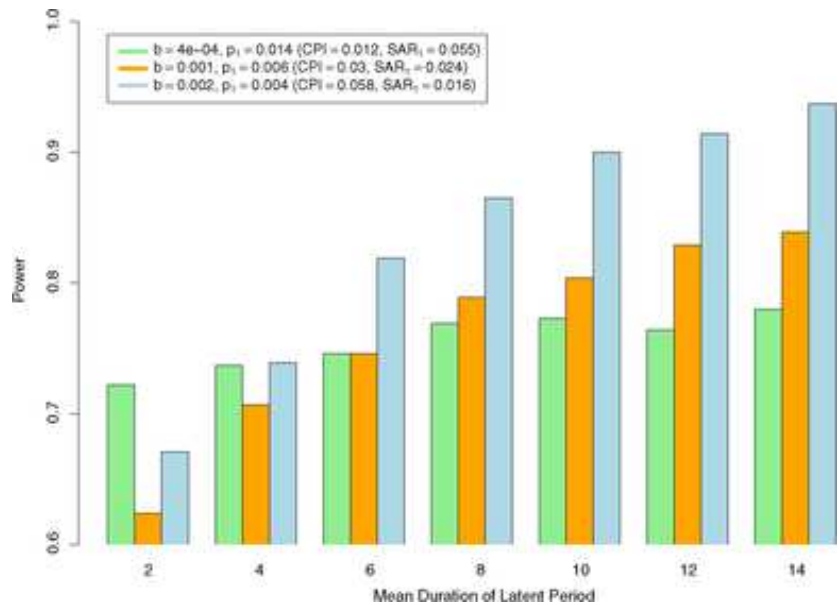

Fig. 5.  *Trend of changes in power as mean duration of the latent period increases for
different settings of b and $p_1$. Distributions of the latent period are uniform over three
days and correctly specified in the models. Results are based on 2000 simulations. 2000
permuted samples were drawn for each permutation test.*

To investigate how sensitive the statistical power of the permutation test is to the distribution of the latent period, we vary the true mean duration from 1.5 to 14 days, while keeping $g(l)$ a uniform distribution over three days. These distributions of the latent period are correctly specified in the models. We expect to see an increase in power, because increasing the latent period is essentially increasing the generation time between successive cases [Fine (2003)]. To look at the trend of changes in power when $b$ is small, medium and large, simulations were done under three parameter settings: ($b = 0.0004$ [CPI $= 0.012$], $p_1 = 0.014$ [SAR$_1 = 0.055$]), ($b = 0.001$ [CPI $= 0.03$], $p_1 = 0.006$ [SAR$_1 = 0.024$]) and ($b = 0.002$ [CPI $= 0.058$], $p_1 = 0.004$ [SAR$_1 = 0.016$]). The values of $p_1$ are chosen to ensure that the initial power is below 0.8 and has the potential of reaching or exceeding 0.8. Results are displayed in Figure 5. Overall, power increases, and the rate of increment decreases, as the mean duration of the latent period (and thus the generation time) becomes longer. However, the rate of increment is higher at larger values of $b$, which means that the power of the refined permutation test is more sensitive to the distribution of the latent period when $b$ is large. Such sensitivity does not compromise the usefulness of the permutation test, since our simulation study is performed under the setting with the minimum level of power. For avian influenza, the mean latent period may be as long as 14 days, and the power will very likely be higher than in our simulation setting.

**4. Discussion.** We have proposed a simple permutation method and its refined version to test the presence of person-to-person transmission within or between households. Using simulations, we have shown that the resampling methods are comparable to or outcompete the standard asymptotic testing method where such asymptotic method is applicable. More importantly, the resampling methods remain valid in many settings where the asymptotic method is not applicable or not available yet. We have shown that, for an infectious disease with relatively rare incidence, person-to-person transmission could still be detected with decent power even if the total number of cases is as few as seven or eight, given that the transmission probability is high and the population is relatively large. We have studied the statistical power of the resampling methods under the model with two levels of contacts: within households and between households. The methods could be generalized to models with additional clustering groups such as schools and work places.

We have assumed that the latent and incubation periods are identical and that the distributions of the latent and infectious periods are known. Other assumptions about the relation between the latent and incubation periods could be made, but may lead to different inference procedures and conclusions. As the presence of the infectious period implies nonzero transmission probabilities, the actual alternative hypothesis we are testing is $p_1 > 0$ or $p_2 > 0$ and $\eta \sim f(l)$, that is, $f(l)$ is a part of the parameters, but we fix it rather than estimate it. Estimating $g(l)$ and $f(l)$ solely from a sequence of symptom onsets is an ongoing research topic and is only practical for a relatively large number of cases [Wallinga (2004), Cauchemez (2006)]. To use our method in real epidemics, one could choose a range of plausible settings of $g(l)$ and $f(l)$, and any setting yielding a significant p-value is a warning sign of transmission between human beings. Appropriate adjustment for multiple testing could be used, but one should be aware that these tests are highly correlated as they are essentially based on the same data set, and a Bonferroni-type adjustment is likely to be over-conservative.

In our simulation study the likelihood is calculated up to day $T - \delta_{max}$ for subjects who do not show symptoms up to day $T$, an incomplete adjustment for right-censoring of infection status. A complete adjustment should take into account that infection might have occurred after $T - \delta_{max}$ and the latent period extends over $T$. Complete adjustments may be important for real-time analysis, especially when $T \gg \delta_{max}$ does not hold. In our simulation setting, $T \gg \delta_{max}$ approximately holds, and the bias in parameter estimates induced by right-censoring is minimal according to the simulation results in Yang, Longini and Halloran (2006).

When conducting the test, maximum likelihood estimates of $b$, $p_1$ and $p_2$ are obtained. From these, estimates of other quantities such as the local reproductive number $R$ and SAR can be derived. We note that, fixed at a value as small as 0.00005 (SAR$_2$ = 0.0002), $p_2$ is generally underestimated due to limited information and, consequently, $R$ is also biased downward. Based on simulation results (not shown), the bias decreases as the true value of $p_2$ or size of the data increases.

We have assumed that each susceptible individual is exposed to an external common infectious source up to day $S$. One may argue that such exposure may only be reasonable for a subset of the population in some situations. Our model can be applied to such situations as well, but only when there is no infected case in the subpopulation which is not exposed to the common source; otherwise, person-to-person transmission exists for sure. In addition, the exposure level to the common source can be assumed as varying from household to household, but permutation should be restricted within households and inference must be supported with sufficient data.

In real epidemics, statistical inference may be very sensitive to the specification of $S$. Particularly, mis-specifying a smaller value for $S$ will likely increase the type I error, as cases that appear after $S + \delta_{max}$ must be accounted for by intensive person-to-person transmission. If no relevant information is available for determining $S$, assuming $S \geq T$ will yield the most conservative p-value. Changing the value of $S$ may affect the admissibility of models, depending on the specification of $g(l)$ and $f(l)$. To apply our methods, it is necessary to ensure that both the null and the alternative models are admissible under these assumptions. Additionally, it may be difficult to identify a clear cut point for the common source exposure, and how to impose the censoring mechanism on $S$ without compromising the test performance is open to further research.

Early detection of person-to-person transmission from limited data is crucial in containing pandemics of emerging infectious diseases such as avian influenza, and our work provides an effective tool for such evaluation. Our method requires not only a time sequence of symptom onsets, but also data on membership of households, whether or not they have cases. We believe that such data requirements are reasonable, and that the information could be collected by local health authorities. When only households with cases are available, selection bias needs to be addressed to make the test valid, which is a topic for further investigation.

## APPENDIX

**A.1. Statistical model.** Assume that the epidemic starts on day 1 and stops by day $T$ in a population of size $N$. Let $\tilde{t}_i$ be the symptom onset day for an infected person $i$. The probability that an infective family member $j$ infects subject $i$ on day $t$, given that subject $i$ is not infected up through day $t - 1$, is expressed as

$$（2） \qquad p_{ji}(t) = p_1^{I(j \in H_i)} p_2^{I(j \notin H_i)} f(t - \tilde{t}_j),$$

where $I(\cdot)$ is the indicator function (1: true, 0: false), $H_i$ is the set of people in the same household with person $i$, and $f(l)$ is the distribution of the infectious period. The probability that subject $i$ escapes infection from all infective sources on day $t$, conditioning on that subject $i$ is not infected up through day $t - 1$, is then given by

$$（3） \qquad e_i(t) = (1 - b)^{I(t \leq S)} \prod_{j=1}^{N} p_{ji}.$$

Because the exact infection date is unobservable, we assume that the duration of the latent period $\delta$ is distributed as $g(l) = \Pr(\delta = l)$, $l = \delta_{\min}, \delta_{\min} + 1, \ldots, \delta_{\max}$, so that we can construct a likelihood for person $i$ as the following:

$$（4） \qquad L_i(b, p_1, p_2 | \tilde{t}_j, j = 1, \ldots, N)$$

$$= \begin{cases} \displaystyle\prod_{t=1}^{T} e_i(t), & \text{not infected,} \\[2ex] \displaystyle\sum_t g(\tilde{t}_i - t)(1 - e_i(t)) \prod_{\tau=1}^{t-1} e_i(\tau), & \text{otherwise.} \end{cases}$$

The overall likelihood $L(b, p_1, p_2|\tilde{t}_i, i = 1, \ldots, N) = \prod_i L_i(b, p_1, p_2|\tilde{t}_j, j = 1, \ldots, N)$ for the full model is maximized with respect to $b$, $p_1$ and $p_2$ to obtain the MLEs of the three parameters, and from these, the estimates of CPI, SARs and $R$. For notational convenience, we suppress the information about household membership that should appear behind the condition symbol in $L$. When there is no person-to-person transmission, that is, $p_1 = p_2 = 0$, (3) reduces to

$$e_i(t) = (1 - b)^{I(t \le S)}.$$

Let $L_0(b|\tilde{t}_i, i = 1, \ldots, N)$ denote the likelihood for the null model. The test statistic is defined as in (1).

## A.2. Null distribution.

A.2.1. *Resampling distribution.* Consider the observed data set as a sample point from the space of all possible infection status and symptom onset times that could occur based on the given population and parameter setting. There exists a class of sample points, which we refer to as the likelihood equivalence class, that have the same likelihood $L_0(b|\tilde{t}_i, i = 1, \ldots, N)$ as the observed data under the null hypothesis $\mathcal{H}_0 : p_1 = p_2 = 0$. If the null hypothesis is true, each sample point in the class occurs with equal probability. That is, if such a class is identifiable, we can obtain the null distribution of the test statistic by resampling sample points from the class with equal probability. Clearly, sample points obtained by permuting the observed infection status and associated symptom onset dates across the population belong to the likelihood equivalence class. Generally, the whole likelihood equivalence class is difficult to identify, and the use of permuted samples is straightforward and fruitful. Let $(\tilde{t}_1^{[k]}, \tilde{t}_2^{[k]}, \ldots, \tilde{t}_N^{[k]})$ be the $k$th permuted sample of $(\tilde{t}_1, \tilde{t}_2, \ldots, \tilde{t}_N)$, and let $\lambda^{[k]}$ be the corresponding test statistic, $k = 1, \ldots, M$. Then the empirical distribution of $\lambda^{[k]}$ over all $k$ can serve as the null distribution of $\lambda$, and the p-value is given by $\frac{1}{M} \sum_k I(\lambda \ge \lambda^{[k]})$.

In our situation, however, it is possible to identify a subset of the likelihood equivalence class which is much larger than and that contains the permuted samples. The idea is more clearly illustrated in the situation without the latent period. Suppose that infection times are observable, and let $\tilde{t}_i$ denote the infection time instead of the symptom onset time for now. Then, the likelihood for the null model is given by

$$
\begin{aligned}
L_0(b|\tilde{t}_i, i = 1, \ldots, N) &= \prod_{i \in \overline{D}} (1 - b)^S \times \prod_{i \in D} \{(1 - b)^{\tilde{t}_i - 1} b\} \\
&= (1 - b)^{(N - \bar{N})S - \bar{N} + \sum_{i \in D} \tilde{t}_i} b^{\bar{N}},
\end{aligned}
$$
(5)

where $D$ is the set of $\tilde{N}$ infected subjects and $\overline{D}$ the set of uninfected subjects. Therefore, one can randomly re-arrange the infection status and infection times while keeping the likelihood value unchanged, as long as the sum of infection times ($\sum_{i \in D} \tilde{t}_i$) and the number of infections ($\tilde{N}$) remain the same. Each re-arrangement is a sample point in the likelihood equivalence class. To keep $\tilde{N}$ unchanged, a permutation of the infection and associated symptom status across the population would suffice, and we refer to it as the initial stage of the resampling procedure. The next stage, which we call the refinement stage, is to draw a sample point with equal probability from all possible distinct re-arrangements of infection times, given the infected cases are fixed. If the refinement stage is not carefully planned, the principle of equal probability can be easily violated, and the consequence is incorrect type I error and/or insufficient statistical power. The problem can be re-stated as sampling with equal probability from all distinct arrangements of $n$ balls (sum of infection times) into $m$ boxes (infected cases), each box with a fixed volume of $v$ ($S$). Let $W(n, m, v)$ be the number of all possible distinct arrangements for such condition. This is a recursive system that can be solved by

$$(6) \qquad W(n, m, v) = \sum_{k=0}^{\min(n,v)} W(n - k, m - 1, v),$$

with the stopping rules $W(n, 0, v) = 0$, $W(0, m, v) = 1$ and $W(n, 1, v) = I(n \leq v)$. An arrangement can be sampled with equal probability through the following procedure:

1. Start with the box labeled $i = 1$, and there are $N_1 = n$ balls to be distributed.
2. In step $i$, let $N_i$ be the number of balls not distributed yet. Randomly choose an integer $n_i$ from $(0, 1, \ldots, r)$ according to the weights $W(N_i - k, m - i, v)$, $k = 0, 1, \ldots, r$, where $r = \min(N_i, v)$, and assign $n_i$ balls to box $i$. Let $N_{i+1} = N_i - n_i$, and go to box $i + 1$.
3. In the last step, distribute all the remaining $N_m$ balls to box $m$.

$N_m$ will not exceed $v$ for sure, because in step $m - 1$ any arrangement resulting in $N_m > v$ has a weight of 0 and thus is excluded from sampling. Hence, this sampling procedure has the advantage of looping over all boxes only once.

This sampling scheme can be adapted to situations with a latent period, but symptom onset times instead of infection times are subject to re-arrangements. The main deviation from the above ideal situation is that, because some cases may have special exposure history, re-arrangement of their symptom onset times will likely change the whole likelihood, and thus, they should be excluded from the refinement stage. One example is seen in simulations, where we let the exposure to a common source of infection last from day 1 to day $S$, and let the latent period vary from $\delta_{\min}$ to $\delta_{\max}$ days. For any case $i$ with symptom onset time $\tilde{t}_i > \delta_{\max}$, there are $\delta_{\max} - \delta_{\min} + 1$ days in which infection could happen, that is, any day between $\tilde{t}_i - \delta_{\max}$ and $\tilde{t}_i - \delta_{\min}$. Symptom onset time of case $i$ could be re-arranged from day $\delta_{\max} + 1$ to day $S + \delta_{\min}$ without changing the likelihood of the null model, as long as the sum of symptom onset times are not changed. However, there may be cases with symptom onset between day $\delta_{\min} + 1$ and day $\delta_{\max}$, for whom the number of days in which infection could happen is less than $\delta_{\max} - \delta_{\min} + 1$. Re-arrangement of symptom onset times of these cases will very likely change the likelihood because the number of potential infection days will also change. Similarly, cases with symptom onset after day $S + \delta_{\min}$ should be excluded from the refinement stage as well.

A.2.2. *Asymptotic distribution.* While the asymptotic null distribution of $\lambda$ is not readily available for testing $\mathcal{H}_0 : p_1 = p_2 = 0$, it is available for testing $\mathcal{H}_0 : p_1 = 0$ if we fix $p_2 = 0$, that is, infection is only possible by the common source or within-household contacts. In this two-parameter setting, the escape probability for person $i$ on day $t$ given the existence of person-to-person transmission is

$$e_i(t) = (1-b)^{I(t \leq S)} \prod_{j \in H_i} (1 - p_1 f(t - \tilde{t}_j)),$$

and the test statistic is

(7)                 $$\lambda = -2 \log \frac{\sup_b L_0(b | \tilde{t}_i, i = 1, \ldots, N)}{\sup_{b, p_1} L(b, p_1 | \tilde{t}_i, i = 1, \ldots, N)}.$$

Self and Liang ([1987](#)) showed that $\lambda \sim \frac{1}{2} \chi_0^2 + \frac{1}{2} \chi_1^2$ under $\mathcal{H}_0 : p_1 = 0$ in such a model, where $\chi_0^2$ is constant 0 and $\chi_1^2$ is a Chi-square random variable with one degree of freedom.

**A.3. Simulation study in the two-parameter setting.** We compare the resampling test to the asymptotic test via a simulation study for the two-parameter setting. Only data observed up to day $S$, the last day of exposure to the common infective source, are used for testing to make the comparison fair, because the asymptotic test cannot handle data beyond day $S + \delta_{\max}$. The resampling method has two variations, one involving only the initial permutation stage, and the other having both stages. The former is referred to as the simple permutation test, which is widely applied to many problems; and the latter is called the refined permutation test in this paper to make a distinction between these two variations. We shall show through simulations that the refined permutation test has some advantages over both the simple permutation test and the asymptotic test for small sample sizes, and that *Comparison of type* I *error and power between the permutation test and the asymptotic test for models with only b and $p_1$. The community is composed of 4 households of size 5. Results are based on 5000 simulations. 2000 permuted samples were drawn for each test*

| $b$ | CPI | $p_1$[a] | $SAR_1$ | $N_{idx}$[b] | $N_{tot}$[c] | Asymptotic | Simple permutation | Refined permutation |
|-----|-----|-----|-----|-----|-----|-----|-----|-----|
| 0.01 | 0.26 | 0.0 | 0.0 | 3 | 5 | 0.029 | 0.039 | 0.050 |
| | | 0.02 | 0.078 | 3 | 6 | 0.21 | 0.22 | 0.26 |
| | | 0.05 | 0.18 | 3 | 8 | 0.60 | 0.57 | 0.63 |
| | | 0.08 | 0.28 | 3 | 10 | 0.85 | 0.81 | 0.85 |
| 0.02 | 0.45 | 0.0 | 0.0 | 4 | 9 | 0.034 | 0.046 | 0.049 |
| | | 0.02 | 0.078 | 4 | 10 | 0.21 | 0.21 | 0.24 |
| | | 0.05 | 0.18 | 4 | 12 | 0.60 | 0.54 | 0.63 |
| | | 0.08 | 0.28 | 4 | 14 | 0.87 | 0.79 | 0.87 |
| 0.03 | 0.6 | 0.0 | 0.0 | 4 | 11 | 0.048 | 0.049 | 0.048 |
| | | 0.02 | 0.078 | 4 | 13 | 0.18 | 0.19 | 0.22 |
| | | 0.05 | 0.18 | 4 | 15 | 0.55 | 0.48 | 0.58 |
| | | 0.08 | 0.28 | 4 | 16 | 0.80 | 0.67 | 0.81 |

[a]Type I errors are reported when $p_1 = 0$.
[b]$N_{idx}$ is the average number of index cases.
[c]$N_{tot}$ is the average total number of cases.

the three tests tend to be equivalent for large sample sizes. By large sample size, we mean both a relatively large population and a large number of cases of the disease.

We first present simulation results in Table 2 for a small population composed of 4 households, each of size 5. Values of $b$ and $p_1$ are chosen to cover a full range of statistical power levels. When $p_1 = 0$, the reported values are type I errors. Clearly, the refined permutation test preserves type I error at the specified level of 0.05 for all settings of $b$. The asymptotic test is the most conservative in rejecting the true null hypothesis by having the smallest type I errors when there are 10 or fewer cases. Surprisingly, the simple permutation test is also conservative when there are only few cases, but less so than the asymptotic test. When $b$ is as large as 0.03 (CPI = 0.6), all methods preserve type I error equally well. In terms of statistical power, the refined permutation test is superior to both of the other two methods. The simple permutation test, however, has the lowest power when there is a fair number of secondary (nonindex) cases, especially when both $b$ and $p_1$ are large.

In Table 3 the population size is increased to 500 with 100 households. Similar to Table 3, we observe that the asymptotic test is conservative with the type I errors much lower than 0.05. When $p_1$ is relatively small, that is, at the second row for each level of $b$, the asymptotic test is not as powerful as the resampling methods. The three methods tend to have the same performance

Table 3

*Comparison of type* I *error and power between the permutation test and the asymptotic test for models with only b and $p_1$. The community is composed of 100 households of size 5. Results are based on 2000 simulations. 2000 permuted samples were drawn for each test*

| $b$ | CPI | $p_1$[a] | $SAR_1$ | $N_{idx}$[b] | $N_{tot}$[c] | Asymptotic | Simple permutation | Refined permutation |
|---|---|---|---|---|---|---|---|---|
| 0.0005 | 0.015 | 0.0 | 0.0 | 7 | 7 | 0.037 | 0.042 | 0.046 |
| | | 0.010 | 0.039 | 7 | 8 | 0.51 | 0.52 | 0.53 |
| | | 0.020 | 0.078 | 7 | 9 | 0.78 | 0.77 | 0.78 |
| | | 0.030 | 0.11 | 7 | 10 | 0.87 | 0.86 | 0.87 |
| 0.0010 | 0.03 | 0.0 | 0.0 | 13 | 14 | 0.031 | 0.047 | 0.047 |
| | | 0.010 | 0.039 | 13 | 16 | 0.59 | 0.64 | 0.64 |
| | | 0.015 | 0.059 | 13 | 17 | 0.78 | 0.81 | 0.81 |
| | | 0.020 | 0.078 | 13 | 18 | 0.88 | 0.90 | 0.90 |
| 0.0050 | 0.14 | 0.0 | 0.0 | 51 | 66 | 0.037 | 0.049 | 0.053 |
| | | 0.005 | 0.020 | 51 | 69 | 0.43 | 0.45 | 0.47 |
| | | 0.010 | 0.039 | 51 | 74 | 0.85 | 0.85 | 0.86 |
| | | 0.015 | 0.059 | 51 | 78 | 0.97 | 0.97 | 0.97 |

[a]Type I errors are reported when $p_1 = 0$.
[b]$N_{idx}$ is the average number of index cases.
[c]$N_{tot}$ is the average total number of cases.

when $p_1$ increases. Again, the refined permutation method seems to be the best choice in these circumstances.

Y. YANG
I. M. LONGINI
M. E. HALLORAN
PROGRAM OF BIOSTATISTICS AND BIOMATHEMATICS
DIVISION OF PUBLIC HEALTH SCIENCES
FRED HUTCHINSON CANCER RESEARCH CENTER
SEATTLE, WASHINGTON 98109-1024
USA
E-MAIL: yang@scharp.org
        ira@scharp.org
        betz@scharp.org